\documentclass{article}
\usepackage{graphicx}
\usepackage{amssymb, amsmath}
\usepackage{amsfonts}
\usepackage{epstopdf}

\title{How Stands Collapse I}
\author{Philip Pearle\\ 
Hamilton College\\
 Clinton, NY 13323, USA\\
e-mail: ppearle@hamilton.edu}

\begin{document}
\date{}
\maketitle
\abstract{In this volume in honor of GianCarlo Ghirardi, I discuss my involvement with ideas of dynamical collapse of the state vector.  10 problems are introduced, 9 of which were seen following my initial work.  4 of these problems had a resolution in GianCarlo Ghirardi, Alberto Rimini and Tullio Weber's  Spontaneous Localization (SL) model (which added 1 more problem).  This stimulated a (somewhat different) resolution of these 5 problems in the Continuous Spontaneous Localization (CSL) model, in which I combined my initial work with SL.  In an upcoming  volume in honor of Abner Shimony I shall discuss the status of the 5 remaining post-CSL problems.}

\section{How I Got Into This.}
	\textit{\qquad\qquad\qquad\qquad \qquad \qquad\qquad To be, or not to be: that is the question.}\\
 
 	The editors' charge is to ``... share with the widest possible community your views on quantum theory ....  with the emphasis on conceptual developments, both those achieved and those hoped for," and  to give a ``...  personal assessment of the role of quantum theory in your work ... ."  The word ``role"  brought to mind the role of Hamlet when, in  Act III, Scene I, he delivers the lines quoted above.  For, my professional life has mostly been devoted to going beyond standard quantum theory, which describes what might be, to construct a theory which separates what is to be from what is not to be, i.e., a quantum theory of reality.    
	
	The personal and the professional are highly intertwined.  Although the former is usually expunged from papers, I imagine it need not be so here\cite{AbnerFest}.  Like most of us, my first encounter with quantum theory was in college, but for me it was a career-changing experience.  I was in a cooperative program in electrical engineering, which involved alternately working at Bell labs and M.I.T. until I got two degrees and could, presumably, go out and earn a living.  While I was at college, some of my best friends were physics majors, inclined toward theoretical physics.  When their conversation touched quantum theory, I felt left out of some arcane, mysterious realm.  I therefore enrolled in the introductory course in quantum theory.  It was taught by Felix Villars, and our textbook was Bohm's\cite{Bohmtext}.  
	
	I didn't understand very well what was going on in either the lectures or the book.  Why was there so much stuff about waves when we were talking about particles?  In electrical engineering, we knew the difference between waves and particles.  I kept waiting for the teacher or the book to tell me where the particles are.  Halfway through the course, when we got to the harmonic oscillator,  it dawned on me that they weren't ever going to tell me where the particles are. I got upset. 
	
	I had never before gone to a teacher's office hours, because I could figure things out for myself.  But, I had thought as hard as I could, and I had come up with a question which I could not answer.  In classical mechanics, given the  particles' initial positions and velocities, you could use Newton's second law to predict  them forever after.  Well, apparently Schr\"odinger's equation was the replacement for Newton's second law since, given the initial wave function you could use Schr\"odinger's equation to predict it forever after. I could see how one could get particle initial conditions: they are ``there."  But, the wave function is a complicated thing, with lots of ins and outs, and complex to boot, and it isn't ``there."  So I went into Professor Villar's office.  No one else was there.  I asked him my blockbuster question:  ``Where do you get the initial wave function?"  And his  answer was immediate: ``From the previous wave function."
	
	My instantaneous reaction was, that isn't right:  it was as if he had said that you never needed initial conditions in mechanics.  But, he immediately qualified it, for what he meant was ``From the wave function which resulted from the outcome of the previous experiment."  I do not now remember much of what he said thereafter, but I remember feeling, as I left, that my question had been answered, I guess, but I felt a disquiet.  
	
	Only after reflection was I able to articulate it to myself: instead of a one-time application of initial conditions, it appeared that they had to be applied endlessly.  It seemed one could not do quantum theory unless repeatedly interrupted by an experimenter bursting in with the latest news.  Part of the charm of theoretical physics to me was that it had the aura of a heroic, independent enterprise.  One could do it alone, with just a pencil and paper, much like my dad, who was a CPA, and who, many evenings, would sit at a card table surrounded by lots of papers, beautifully writing important numbers in neat columns in large lined books.  But, every day, he had to go out to get information from businessmen.  Apparently, doing quantum physics was just like doing accounting: one needed a pencil and paper and information from experimenters.   
		
	 I learned to play the game in order to pass the tests, but I was not happy with the subject.   I couldn't believe that this was the best that could be done, that there wasn't more to it than I was being taught.   However, one day, near the end of the semester, Professor Villars told us something which, as much as anything else, put me on the road to my life's work.  He said that there was a section at the end of the book which he wasn't going to assign but which we could read if we were interested.  In it, Bohm describes an argument due to Einstein, that quantum theory isn't the last word, but which Bohm then refutes. 
	 
	 That put me instantly on alert. To a Jewish boy growing up in the South Bronx before, during and after World War II, Einstein was a revered figure.  When my mom proudly showed my grade school report card to my Hungarian grandma who lived across the courtyard, she would pat me on the head and pronounce, ``A leetle Einshtein."  At a later age, around the block, when playing slug, stoop ball, or stick ball with a ``spaldeen" (a pink ball made by the Spalding company), if by dint of one's personal efforts the ball got lost down the sewer, some friend was sure to say ironically, ``Hey, Einstein."  So, I felt excited: I had independently arrived at an opinion of quantum theory which turned out to be the opinion of the great Einstein.  
	 
	 However,  I also had an inordinate respect for the authors of textbooks. For example, the authors of history textbooks explained the mistakes and bad behavior in the past so clearly that I was sure they would not be repeated, and I had found it comforting that the world therefore had to be getting better. And, the authors of technical textbooks  
were right, as I knew from electrical engineering, where we checked out everything they said in innumerable labs.    So, into my mind immediately flashed a conflict.  

	But, the next thing Professor Villars said was that, although in the textbook Bohm had said that Einstein was wrong, after he had stopped writing the book, Bohm had said that Einstein was right.  He had then gone on to write papers in which something is revealed behind quantum theory (Hamlet, Act I, Scene V:  ``There are more things in heaven and earth, Horatio, than are dreamt of in your philosophy"). So, the conflict was resolved: both Einstein and the textbook author were of my opinion.
	
	I went off and read the textbook on the EPR paradox, and read Bohm's papers\cite{Bohmpapers}.   I did not understand either very well.  But, I did get this much: I had concluded that no one was ever going to tell me where the particles were and---there were the particles!  
	
	When I finished my engineering degrees, I decided to go for a PhD in physics, in large part because I wanted to see what was behind quantum theory.  I stayed at M.I.T. but, just as in the electrical engineering department, where I was very interested to learn how radios work and, in five years, could never take a course to tell me that, in the physics department  there was never a course  which taught  what I wanted to know.  When I mentioned my interest to my prospective thesis advisor, he changed the subject.  So, having taken quantum field theory courses from Professors Silvan Schweber and Kenneth Johnson, all I was fit for was to do something in particle physics. After getting my degree, I went to teach at Harvard but, after trying, I could not work up enthusiasm for Regge poles and bootstraps.
	  		
	Even back in the early 60's, the qubit played a role in the field of foundations of quantum theory ---most universities had 0 people interested in the subject, but occasionally there was 1.  At Harvard, I found Wendell Furry and, with his encouragement, wrote my first paper entitled "Elimination of the Reduction Postulate from Quantum Theory and A Framework for Hidden Variable Theories."  It was really two papers.  
	
	The first argued that the reduction (collapse) postulate of the Copenhagen school is ill-defined, since it never precisely specifies \textit{when} and \textit{why} the state vector should be reduced.  However, it must be reduced (e.g., following a completed experiment)  if one interprets the state vector as corresponding to reality, since a  superposition of macroscopic states does not correspond to reality.   One can eliminate the ill-defined collapse postulate if one gives up the reality interpretation, and replaces it by the ensemble interpretation (that the state vector corresponds to an ensemble), Einstein's favored interpretation.  (I did not appreciate then, as I do now, that the \textit{which},  the states which make up the ensemble, are generally ill-defined, no less than the collapse postulate's end states). But, the ensemble interpretation entails that quantum theory does not describe individual objects and, since we know that individual objects exist, this encourages looking for a (so-called hidden variables) theory which does describe individual objects.
	
	So, the second argument set out an abstract framework for hidden variables theories.  I gave an example of  that framework for a two-state quantum system.  (The hidden variable space is the surface of a sphere: when a spin state vector points in the direction $\hat {n}$, this means a physical system is represented by a hidden variable point $\hat {r}$ on the upper hemisphere with $\hat {n}$ as its ``north pole," and that point is occupied with probability $\hat {n}\cdot\hat {r}/\pi$.  When the spin is measured in a direction $\hat {m}$, the outcome will be ``up" if the hidden variable point lies in the upper hemisphere with $\hat {m}$ as its north pole, and ``down" if  the hidden variable point lies in the lower hemisphere). My problem was that, for the life of me, I couldn't find an example which fitted my framework for a three-state (or higher) quantum system.   
	
	I sent the preprint to various people, and was excited to receive replies from John Bell and Eugene Wigner!  Bell suggested that I look at a paper by Gleason\cite{Gleason}, so I learned that I was never going to find a higher state generalization of my framework (in current language, my framework demanded non-contextuality). So, I dropped the second part of the paper and enlarged the first half, taking into account questions raised by Wigner, and called it ``Alternative to the Orthodox Interpretation of Quantum Theory."\cite{PearleAlternative}.  	
	
	(I later heard that eventually someone else found my two-state model.  This experience with a hidden variables model based upon the surface of a sphere led me to  introduce what is now called the ``detection efficiency loophole,"\cite{Pearleloophole}, based upon a hidden variables model  whose hidden variables space is the interior of a sphere. Here, a way around Bell's inequality is suggested if the pair of spin-1/2 particles which reach rather inefficient detectors have three choices, ``spin up,"  ``spin down," and ``undetected," and the ``undetected" data is discarded.)
	
\section{Dynamical Wavefunction Collapse}	
	
	There was then a conflict in my mind.  On the one hand, because my elegant framework for a hidden variables theory could not work, I was down on hidden variables theories, even down on the deBroglie-Bohm pilot wave model which had excited me so much. I thought that quantum theory was more elegant and general than the hidden variables theories I had seen or could imagine.  On the other hand, while I believed that the ill-defined collapse postulate was the Achilles heel of quantum theory, I came to doubt the alternative I had argued for, the ensemble interpretation, because I thought that quantum theory worked so well it ought to be a description of individual reality. 
			
		I started to think about whether one could make collapse well-defined, when a paper by Bohm and Bub\cite{BohmBub} came out, ``A Proposed  Solution of the Measurement Problem in Quantum Mechanics by a Hidden Variable Theory."  In it, the authors provided a dynamical equation to describe state vector collapse.  The final state of collapse is determined by Wiener-Siegel\cite{Wiener-Siegel} hidden variables (essentially, the hidden variables are vectors in a Hilbert space like the one in which the state vector resides).   
				
		The idea of replacing the collapse postulate by collapse dynamics was just the thing I was looking for.  I knew of the Wiener-Siegel hidden variables but, although elegant, they didn't satisfy my framework and I didn't believe they really existed either. I thought that, instead of hidden variables, something random in the Schr\'odinger equation would be more appropriate to determine the final state of collapse, since many things exist in nature which behave in random fashion.
						
		I thought of attacking the problem, of finding an appropriate collapse dynamics, in three stages.   The first stage was to find minimal conditions, on the probability which describes the ensemble of collapsing state vectors, sufficient to entail collapse. The second was to find an evolution equation, for this probability, which satisfies the minimal conditions.  The third, and hardest, stage was to modify Schr\"odinger's equation by adding something random, to thereby obtain an ensemble of collapsing state vectors, whose probability would obey the evolution equation.   	
		
		The first stage turned out simply, which was encouraging.   Idealized, a collapse evolution is
\begin{equation}\label{1}
|\psi,0\rangle=\sum_{n=1}^{N}c_{n}(0) |a_{n} \rangle\rightarrow |\psi,t\rangle=\sum_{n=1}^{N}c_{n}(t) |a_{n} \rangle\rightarrow |\psi,\infty\rangle=c_{m}(\infty)|a_{m} \rangle.  
\end{equation}
\noindent In Eq. (\ref{1}), $|a_{n}\rangle$ is an orthonormal \textit{preferred basis}, to one of which states, $|a_{m}\rangle$,  the state vector eventually collapses and  $|c_{m}(\infty)|^{2}=1$: this end result is to occur for a fraction  $|c_{m}(0)|^{2}$ of the evolutions.  

	If one defines $x_{m}(t)\equiv |c_{m}(t)|^{2}$, the probability density $P(x_{1}...x_{N}, t)$ of the values $x_{1}...x_{N}$ occurring at time $t$ can be used to characterize the behavior of the ensemble.  Defining  
\[
\overline{f(x_{1}(t)...x_{N}(t))}\equiv\int_{0}^{1}dx_{1}...dx_{N} f(x_{1}...x_{N})P(x_{1}...x_{N}, t),
\]
collapse behavior is ensured if 
\begin{subequations}\label{2}
\begin{eqnarray}
P(x_{1}...x_{N}, t)&\sim&\delta(x_{1}...+x_{N}-1),\label{2a}\\
\overline {x_{n}(t)}&=& x_{n}(0)\qquad\hbox{Martingale property},\label{2b}\\
\lim_{t\rightarrow\infty}\overline{x_{m}(t) x_{n}(t)}&\rightarrow& 0\qquad\hbox{for } n\neq m \label{2c}. 
\end{eqnarray} 
\end{subequations}
\noindent To see this, first note from (\ref{2c})  that the only way the integral of $x_{m}x_{n}P\geq 0$ can vanish is if $P(x_{1}...x_{N}, \infty)$ has delta function behavior where $x_{m}=0$ or $x_{n}=0$ or both $=0$. Since this is true for all $m\neq n$, 
\[
P(x_{1}...x_{N}, \infty)=\sum_{n=1}^{N}K_{n}\delta(x_{1})\delta(x_{2})...\delta(x_{n}-1)...\delta(x_{N}), 
\] 
\noindent where the $K_{n}$'s are constants and (\ref{2a}) has been used.  Finally, applying this to (\ref{2b}) at $t=\infty$ gives the constants: $K_{n}= x_{n}(0)$.  Thus, at $t=\infty$, there are $N$ possible outcomes where all but one of the $x$'s vanishes.  The non-vanishing $x_{n}(\infty)=1$ occurs with probability $x_{n}(0)$: that's collapse behavior. 
 	
	The next stage turned out to be pretty simple too.  Because all $x_{n}$'s must be treated symmetrically, it suffices to start with a two-state system and generalize the result.  The idea is to look for a Fokker -Planck  equation which gives the detailed behavior of  $P(x_{1}, x_{2}, t)$:  
\[
\frac{\partial P(x_{1}, x_{2}, t)}{\partial t}=-\sum_{n=1}^{2}\frac{\partial v_{n}(x_{1}, x_{2})P}{\partial x_{n}}+
\sum_{m,n=1}^{2}\frac{\partial^{2} A_{nm}(x_{1}, x_{2})P}{\partial x_{n}\partial x_{m}},
\]	
\noindent  and find the quantities $v_{n}$, $A_{nm}$ (a symmetric matrix with non-negative eigenvalues) which entail the properties (\ref{2}).  By multiplying by $x_{n}$ and integrating over the square  $0^{-}\leq x_{1}, x_{2}\leq 1^{+}$ (actually, by (\ref{2a}), $P$ only has support on the line $x_{1}+ x_{2}=1$ within this square) one obtains, following integrations by parts, 
\[
\frac{\partial \overline{x_{n}(t)}}{\partial t}=\overline{v_{n}(t)}.   
\]
\noindent	Thus, $v_{n}=0$ in order to satisfy (\ref{2b}) with arbitrary $P$.  

	The condition (\ref{2a}) requires $P(x_{1}, x_{2}, t)=\delta(x_{1}+x_{2}-1)p(x_{1}, x_{2}, t)$ to be a solution, and vanishing of the terms involving second and first derivatives of the delta functions results in the form 
\begin{equation}\label{3}
\frac{\partial p(x_{1}, x_{2}, t)}{\partial t}=\bigg(\frac{\partial}{\partial x_{1}}-\frac{\partial}{\partial  x_{2}}\bigg)^{2}A(x_{1}, x_{2})p(x_{1}, x_{2}, t),
\end{equation}
\noindent in which one is free to set $A(x_{1}, x_{2})=A(x_{1}, 1-x_{1})\equiv A(x_{1})\geq 0$ and $p(x_{1}, x_{2}, t)=p(x_{1}, 1-x_{1},t)\equiv p(x_{1},t)$, so (\ref{3}) is equivalent to 
\begin{equation}\label{4}
\frac{\partial p(x_{1}, t)}{\partial t}=\frac{\partial^{2}}{\partial x_{1}^{2}}A(x_{1})p(x_{1}, t).  
\end{equation}	   
\noindent  If $A(x_{1})$ is a constant, (\ref{4}) is the ordinary diffusion equation, for which probability is not limited to the range $0\leq x_{1}\leq1$ but spreads over the whole real line. Thus the equation has to be supplemented by absorbing boundary conditions at $x_{1}=0,1$. However, if one sets  $A(x_{1})=0$ at  $x_{1}=0,1$,   then  (\ref{4}) automatically provides its own absorbing boundary conditions, since the diffusive motion then  vanishes at the boundary.  

	To see how (\ref{2c}) can be fulfilled, multiply  (\ref{4}) by $x_{1}(1-x_{1})$ and integrate by parts over $0^{-}\leq x_{1}\leq 1^{+}$ to obtain
\begin{equation}\label{5}
\frac{d \overline{x_{1}(t)(1-x_{1}(t))}}{d t}=-2\overline{A(x_{1}(t))}.  
\end{equation}
\noindent  For example, if $A(x_{1})=\alpha x_{1}(1-x_{1})$ then, according to (\ref{5}), $\overline{x_{1}(t)x_{2}(t)}=x_{1}(0)x_{2}(0)\exp-2\alpha t$, and  (\ref{2c}) is satisfied.  

	More generally, 
(\ref{5}) implies that  $\overline{x_{n}(t)(1-x_{n}(t))}$ must keep decreasing until such time $T$ when 
 $\overline{A(x_{1}(T))}=0$.   Since  $\overline{x_{n}(t)(1-x_{n}(t))}\geq 0$, this must occur.  But,   $\overline{A(x_{1}(T))}=0$ is only possible if, at $t=T$, the probability concentrates (has all of its non-vanishing measure) where $A(x_{1})=0$.  Thus, if it is assumed that $A(x_{1})$ only vanishes at the boundary $x_{1}=0,1$,  the probability at $T$ must have the form $p(x_{1},T)=K_{1}\delta(x_{1})+K_{2}\delta(1-x_{1})$.  But, this implies  $\overline{x_{1}(T)(1-x_{1}(T))}=\overline{x_{1}(T)x_{2}(T)}=0$, and (\ref{2c}) is satisfied. 

	The $N$-state diffusion equation which generalizes (\ref{3}) is 
\begin{equation}\label{6}
\frac{\partial p(x_{1},... x_{N}, t)}{\partial t}=
\sum_{n,m=1}^{N}\bigg(\frac{\partial}{\partial x_{n}}-\frac{\partial}{\partial x_{m}}\bigg)^{2}A_{nm}(x_{n}, x_{m})p(x_{1},... x_{N}, t),
\end{equation}	
\noindent where $A_{nm}(x_{n}, x_{m})$ only vanishes at $x_{n}=0$, $x_{m}=0$ and is positive elsewhere.

	Now came the hardest task, constructing a modified Schr\"odinger equation with something random in it, whose ensemble of solutions satisfies (\ref{6}).  I thought to use, as the random variables, the initial phase angles of the amplitudes multiplying the states in a macroscopic superposition.  The advantage was that no new variables needed to be added to quantum theory.   I expected that the term added to Schr\"odinger's equation would be nonlinear in the amplitudes, and it would keep roiling the phases so that they evolved in random fashion.  But, I did not know any formalism which, given a guessed nonlinear term, with the assumption of random initial phases, would enable me to find the associated Fokker-Planck equation.  

	At this point, it was time for me to go on my first sabbatical.  I wrote to John Bell, inquiring whether I could spend my year at CERN and, in a letter sent on April 27, 1972,  he wrote back:\\ 
			
	``If you were to come to CERN I think you would find in me a kindred spirit, but you might not find any others.  ...  My own occasional excursions into other fields are tolerated aberrations rather than normal activities."...\\
		
\noindent and suggested that I contact Josef Jauch at the University of Geneva.  So, that was where I spent my sabbatical year.  Both Jauch and Bell were interested and encouraging, but neither they nor anyone else I encountered could help me go further, and I spent my sabbatical working at something else.  

	But, about a year later, while browsing  by chance through the COOP bookstore in Harvard square, I found---on sale!--- a book by Prigogine\cite{Prigogine}, and I later encountered the Dover book which contains a 1943 review by Chandrasekhar\cite{Chandrasekhar}:  I had two different sets of tools.  Using either, in the notation of  Eq. (\ref{1}), I found that the dynamical equation 
\begin{equation}\label{7}
i\frac{dc_{n}(t)}{dt}=\omega_{n} c_{n}+(c_{n}^{*}(t))^{-1}(c_{n}(t))^{r}\sum_{m=1}^{N}\alpha_{nm}(c_{m}^{*}(t))^{r}
\end{equation}
\noindent  (where $\omega_{n}\equiv\langle a_{n}|H|a_{n}\rangle$ and $\alpha_{nm}$ is a Hermitian matrix) leads to Eq. (\ref{6}) with $A_{nm}(x_{n}, x_{m})\sim |\alpha_{nm}|^{2}(x_{n}x_{m})^{r}$.  So, I finally wrote a paper on all this\cite{Pearle1}.  I was fortunate in my referee, Fred Belinfante, who telephoned me to ask some questions, so that he could submit an unquestioningly positive report, for he said that any negative remark would be taken by the editors of Physical Review as an excuse not to publish the paper.  He later sent me an autographed copy of his book\cite{Belinfante}! 

	In those days, it was frequently difficult for work which considered alternatives to quantum theory to receive  fair consideration (this has only  somewhat abated). Paradoxically, physicists who are often radical thinkers, are quite conservative when it comes to questioning the validity of their bread and butter, standard quantum theory. 
		
\section{Stochastic Differential Schr\"odinger Equation}

	The methods of Prigogine and Chandrasekhar were rather complicated, and entailed some assumptions.  I gave up on the random phases, the idea that nothing extra had to be added to quantum theory to explain collapse.  I considered putting a new physical variable, randomly fluctuating but (for now) otherwise unspecified, into the modified Schr\"odinger equation, and went to the shelves in my college's library looking for a way to analyze differential equations with something random in them. I found, to my surprise and delight, two books on stochastic differential equations\cite{McKean, Wong}.  I'd never heard of stochastic d.e.'s before.  
	
	These are equations of the form
\begin{equation}\label{8}
dx_{n}(t)=G_{n}(x_{1},...x_{N},t)dt+\sum_{m=1}^{N}F_{nm}(x_{1},...x_{N},t)dB_{m}(t),
\end{equation}
\noindent where $B_{m}(t)$ are independent Brownian motion functions ($dB_{n}(t)dB_{m}(t)=\sigma^{2}\delta_{nm}dt$) and $dB_{m}(t)/dt=w_{m}(t)$ are independent white noise functions.  I devoured the books.  Working  with stochastic d.e.'s was fun, like doing magic. There are two forms, one due to Stratonovich, the other due to  It\^o, each  readily convertible into the other, differing respectively in the definition of $F(x,t)dB(t)$, so that $\overline{F(x,t)dB(t)}\neq 0$ or $=0$.  Each form has its advantages. The Stratonovich form allows one to integrate equations using the ordinary rules of calculus.  From the It\^o form one can just pluck out the Fokker-Planck equation governing  the  probability describing the ensemble of solutions, i.e.,  Eq. (\ref{8}) implies
\begin{equation}\label{9}
\frac{\partial P(x, t)}{\partial t}=-\sum_{n=1}^{N}\frac{\partial G_{n}P}{\partial x_{n}}+
\frac{\sigma^{2}}{2}\sum_{m,n=1}^{N}\frac{\partial^{2} \sum_{k=1}^{N}F_{nk}F_{mk}P}{\partial x_{n}\partial x_{m}}.
\end{equation}	
\noindent Using this, it was possible to show\cite{Pearle2} that amplitudes obeying 
\begin{equation}\label{10}
i\frac{dc_{n}}{dt}=\omega_{n} c_{n}+\frac{c_{n}}{c_{n}^{*}}\sum_{m=1}^{N}\alpha_{nm}w_{nm}c_{m}^{*},
\end{equation}
\noindent  leads to (\ref{6}) with $A_{nm}(x_{n}, x_{m})=(\sigma^{2}/2)|\alpha_{nm}|^{2}x_{n}x_{m}$  (compare  Eq. (\ref{7}) with $r=1$, which I had come to favor because of its simplicity). 
In Eq. (\ref{10}), $w_{nm}(t)$ is a Hermitian matrix of independent white noise elements and $dB_{mn}(t)dB_{rs}^{*}(t)=\sigma^{2}\delta_{mr}\delta_{ns}dt$. 

	This idea, of adding a stochastic term to the Schr\"odinger equation so that it becomes a	stochastic differential equation, has proved very fruitful, not only to describe collapse, as shall be enlarged upon, but for other purposes as well\cite{AdlerPercival}. 
	
	I soon found a charming analogy for this collapse dynamics\cite{coins}, useful for providing an intuitive and non-technical explanation of how it works. I happened to be browsing in Feller's book on probability\cite{Feller} (a favorite textbook, from an undergraduate course taught by Stanislaus Ulam) when I encountered the gambler's ruin game. Two gamblers, initially possessing, respectively, a fraction $x_{1}(0)$, $x_{2}(0)$ of their combined wealth (so $x_{1}(0)+x_{2}(0)=1$) repeatedly toss a fair coin, and the result, heads or tails, determines which one gives a dollar to the other.  They play until one gambler loses all his money, and the game ends. The analogy is that the amount of money possessed by one gambler at any time is proportional to the squared amplitude of one of two states whose sum is the state vector representing the physical system undergoing collapse.  Just as one gambler loses all his money, so one of the states loses all its amplitude, and as the other gambler wins all the money, so the state vector ends up as totally described by the other state.   
	
	Let $Q(x)$ be the conditional probability that a gambler wins the game, given that he has  the fraction $x$ of the total wealth. If $\Delta$ is the fraction of the total wealth they exchange at each toss (i.e., $\Delta=$\$1/total dollars), the difference equation 
\[
Q(x)=\frac{1}{2}Q(x-\Delta)+\frac{1}{2}Q(x+\Delta)
\]	 		
\noindent expresses that there are two routes to win if one has fractional wealth $x$, namely lose the next toss and drop to $x-\Delta$ but win thereafter, or win the next toss and rise to $x+\Delta$ and win thereafter.  The solution of the difference equation is $Q(x)=Ax+B$, where $A$ and $B$ are constants.  Since $Q(0)=0$ (because you can't win if you have no money) and $Q(1)=1$ (because you have won if you have all the money), then $Q(x)=x$.  

	That is, if one starts with the fraction $x=x(0)$ of the money, one has the probability $x(0)$ of attaining all the money, $x=1$, which is collapse behavior. The game can be modified to have many players, to have $\Delta$ change as the game progresses (e.g., to get smaller as one gambler gets closer to losing, so as to mimic $A_{nm}(x_{n}, x_{m})$'s decrease as its arguments approach zero), etc.  So, one may think of quantum collapse as a gambler's ruin competition among the states in a superposition, to see which final state wins the game. 
	
	 Indeed, this is just random walk with absorbing barriers. In an appropriate limit where the time between tosses get infinitesimal along with $\Delta$ so that it becomes a time-continuous process, it is described by  Eq. (\ref{6}). 

\section{Problems}

	At this point, how did the collapse program stand?  Although I had found a solution to the three-stage problem I had set for myself, truth to tell, my dynamical collapse was as ill-defined as the Copenhagen collapse postulate.  I had vaguely thought when I had begun that the constraints would be so severe that they would point the way to a unique solution.  Instead, the constraints were not so severe, there was left an array of choices to be made, and I had not a physical principle or principles to rely upon to make those choices.  
	
	What should be the choice of the diffusion matrix $A_{nm}(x_{n}, x_{m})$?  I had made the simplest choice, but others were certainly possible and, even then, what were the matrix elements $\alpha_{nm}$ to be?  In \cite{Pearle1} I wrote ``... the operator $A$ must have nonvanishing matrix elements between states that are macroscopically distinguishable, and so probably will be a nonlocal operator. The magnitudes $|\alpha_{nm}|$ along with $\sigma$ determine the reduction rate.  The rate cannot be too fast, or else the usual quantum predictions will be interfered with.  It also cannot be too slow, or else it will predict that a system can be observed in a superposition of macroscopically distinguishable states.  Perhaps the matrix element magnitudes should be small for a microscopic system, but large for a macroscopic system.  Since a macroscopic system may be thought of as composed of many microscopic systems, such a limitation in magnitudes may act as a constraint on the form of the matrix elements."  
In \cite{Pearle2} I wrote ``... we conjecture ... that the magnitude of matrix elements of $A$ depends only upon the macroscopic distinguishability of appropriate position variables characterizing the states. "  But, these are qualitative statements.  I didn't know how to choose $A$. This might be called the  \textit{interaction problem}.   
		
	Collapse models are phenomenological models, justified because they provide an explanation of  the existence of the world around us and the events that take place in it.   In \cite{Pearle2} I went on, ``But, ultimately, a theory such as this needs to be legitimized by being a consequence of a larger theory that has more ties to established physics. In what areas might such a theory arise? ... $A$ ... is a nonlocal long-range interaction between a system and itself.  This carries the connotation of relevance to self-energy considerations, and perhaps gravitational theory.  Indeed, it is an attractive thought that the juncture between general relativity, which describes events but does not describe microscopic behavior, and quantum theory, which describes microscopic behavior but does not describe events, might be an appropriate place to look."  Call this the \textit{legitimization problem}.  

	 What should be the states $|a_{n}\rangle$ which are the end products of collapse?   This same problem is encountered in standard quantum theory, in applying the collapse postulate, so dynamical collapse was no improvement, in this respect.  It was clear that the preferred macroscopic states had to be spatially localized, since that is what we see. However, the preferred basis couldn't be the position basis since that has infinite average energy and it couldn't be achieved anyway since the competing Schr\"odinger part of the evolution would spread a highly localized wave function.  For dynamical  collapse to be an improvement over postulated collapse, the theory should specify the preferred basis.  So, I came to call this the \textit{preferred basis problem}.  
	 
	 Associated with the preferred basis problem is another problem\cite{PearleNYAS}: ``... we point out that dynamical reduction ... can violate conservation laws.  ...  This is a serious problem for quantum theory with a reduction postulate.  For a dynamical reduction theory it raises intriguing possibilities of deeper insights into the nature of the theory---for example, the possibility that the fluctuating medium that causes the reduction may exchange energy and momentum with the quantum system, or the possibility of rules for choosing the preferred basis states so that conservation law violation is minimized."  So, add to the list the \textit{conservation law problem}. 
	 
	 What turns the collapse dynamics on?  I expected that the correct choice of $A$ would do it\cite{Pearle4}: ``This coupling is expected to increase as the states it couples become more macroscopically different: perhaps it is an increasing function of particle separation and particle masses."  My question to Professor Villars should have the answer that, as with classical physics, you only need to apply initial conditions once.  One should be able to start with an initial state vector and let the modified Schr\"odinger equation take over from there, with collapse automatically occurring as needed, the outcome determined by the particular white noise function used in the Schr\"odinger equation.   But, I didn't have the right $A$, so I manually turned up the interaction when needed, and turned it down when not needed.  In a satisfactory theory, that ought to happen automatically. I came to call this the \textit{trigger problem}.
	  	 
	 What experiments can test the predictions of a collapse theory against those of standard quantum theory?  The obvious test is interference. Suppose an object can be put into a superposition of two states, say with equal amplitudes, and that these states then start to play the collapse game. After a while, the amplitudes of the states will not be equal.  If an interference measurement  on the superposed state is performed for an ensemble, the interference pattern will not  have the same contrast as the pattern would have if the amplitudes had remained equal. I spent my second sabbatical in 1981-1982  at Oxford, invited by Roger Penrose who had just written an article arguing that a satisfactory quantum theory of gravity ought to entail collapse\cite{Penrose}.  Roger received an invitation to attend a conference in Perugia in the Spring, honoring the 90th birthday of Louis de Broglie, and suggested that I attend in his stead. Anton Zeilinger attended the conference,  and told me about a recently completed two-slit neutron interference experiment\cite{Zeilinger}.  As might be expected, the data is consistent with standard quantum theory's prediction, to the experimental accuracy, about 1\%.  But, that does not preclude the possibility that another interference experiment, perhaps one of higher accuracy or one which involves interference of a larger object, might not reveal a discrepancy with standard quantum theory's prediction.  Indeed, data from that experiment enables estimation of a lower limit for the neutron's collapse time of only 5sec.\cite{PearleZeilingerneut}.  Can one perform an interference experiment, or some other kind of experiment, to definitively confirm or rule out dynamical collapse? Call this the \textit{experimental problem}.  
	 
	 Also at this conference in Perugia was Nicolas Gisin, who came with his thesis advisor Constantin Piron, whom I had met in Geneva on my first sabbatical.  Nicolas became interested in this line of research, and soon published an example of a stochastic Schr\"odinger equation, in discussing a dynamical implementation of the von Neumann and Luders projection postulates\cite{Gisin1}.  I pointed out that the collapse time in his example takes an infinite amount of time\cite{PearlePRLlett}.  It is the aim of collapse dynamics to provide a state which corresponds to physical reality---a slogan I like is that what you see (in nature) is what you get (from the theory).  Abner Shimony had argued to me that, for a state which is a superposition of macroscopically distinguishable states, even if the amplitude of one state is huge compared to the others, it is philosophically objectionable to say that the physical reality corresponds to  that one state. Only a finite mean collapse time for the ensemble ensures that there are no tails (other smaller amplitude states in the superposition). 	 This situation has come to be called the \textit{tails problem}.  
	 
	 I had seized upon that as a possible physical principle. In this respect, the dynamical equation (\ref{7}) with  $r=1$  (the choice made in (\ref{10})), where the diffusion matrix is proportional to $x$ near $x=0$, is preferable because the mean collapse time is finite\cite{Pearletime}.  In Gisin's example, the diffusion matrix is proportional to $x^{2}$ near $x=0$, such a slow diffusion that the mean collapse time is infinite.     Gisin's reply\cite{GisinPRLlett} acknowledged the tails problem, but introduced a new and important physical principal.  It is possible to construct a given density matrix from various different ensembles of state vectors. As the density matrix then evolves, Gisin remarked that it ought to evolve in precisely the same way no matter how it was initially formed for, if not, by a judicious local choice of a portion of an entangled  state vector, it could be possible to communicate superluminally if the ensemble behavior far away depended upon that choice.  My models did not have the density matrix evolving independently of its initial constitution.  Up to then, I had never even thought about the behavior of the ensemble of state vectors, so fixated was I upon the problem of making sure that each individual state vector collapsed.  I  then showed that, for there to be no superluminal communication in certain experimental situations, the mean collapse time in dynamical collapse models has to be infinite\cite{PearleSuperlum}.  Thus, resolution of this \textit{superluminal problem} must give rise to the tails problem.  	 
	   
	   At the end of my stay in Oxford, John Bell wrote on July 7, 1982: "It is now clear that I will not get to England again while you are there, so we unfortunately miss each one another on this occasion.  Your nonlinear work is the most serious that I know of, and I wish you great success in continuing it.  It is not so much relativistic invariance that I hope for eventually (I suspect it it is impossible) but a theory covering nevertheless somehow the phenomenon of relativistic quantum theory."  So, add the \textit{relativity problem}.  
	   	   
	 In the remainder of this paper, I shall address progress made on resolving four of these problems and one other problem which later arose (see the next section).  The sequel to this paper shall address the remaining five problems.
	 
\section{Spontaneous Localization}  

	In 1986, John Bell sent me a preprint\cite{BellGRW} which introduced me to the ingenious, insightful, and, indeed, courageous, paper of GianCarlo Ghirardi, Alberto Rimini and Tullio Weber\cite{GRW} describing their ``Spontaneous Localization" (SL) model\cite{GhirBassi}.  I shall content myself with giving a qualitative description of the SL model, since it is my expectation that it will be given a full presentation elsewhere in this volume. 
	
	 It is a model which invokes instantaneous collapse but, unlike the ill-defined Copenhagen postulate of instantaneous collapse which depends upon the imprecise notion of a ``measurement," it gives a precise prescription.  At random times (with suggested rate $\lambda\approx10^{-16}$sec$^{-1}$), a normalized  wave function describing matter containing $N$ particles (electrons and nucleons) is suddenly multiplied  by a gaussian 
$\sim \exp-[({\bf x_{n}}-{\bf z})/2a^{2}]$, and then normalized back to 1, where $\bf{x_{n}}$ is the coordinate of one of the $N$ particles, $\bf{z}$ is randomly chosen according to a certain rule (see below), and the suggested width $a\approx 10^{-5}$cm.  I came to call this multiplication a ``hit," rather than a ``spontaneous localization," thereby saving a mouthful of syllables.  Each particle is independently hit at the same rate $\lambda$.  $\bf{z}$ may be called the center of the hit. The probability of a hit taking place with center at $\bf{z}$ is proportional to the integral of the squared magnitude of the wave function after the hit (but before re-normalization). Thus the hit centers are most likely located where the wave function is largest.   
	 
	 A single particle in a state which is the superposition of two spatially separated wave packets (separation much more than $a$) will, after a hit, most  likely be in a state which corresponds to one of the packets (although altered in shape, and altered in spread if its spread had been larger than $a$), albeit with a small tail of the other packet.  The rate of hits for the single particle is so slow that such single-particle or few-particle superpositions as occur in laboratory experiments will be seldom (and, so, unmeasurably) affected.  However, for a macroscopic object in a state of superposed positions, because of the entangled nature of such a wave function, one hit on one particle suffices to collapse the whole object's wavefunction to one of those positions, albeit with a small tail of the other position. Since all particles experience hits at the same rate, the collapse rate is $N\lambda$, which can be quite rapid for an object with a sufficient number of particles.  
	
	The SL model resolves the interaction, trigger, preferred basis and superluminal problems.   For the trigger problem, the collapse mechanism is always working and automatically is slow for small objects and fast for large ones.  For the preferred basis problem, the preferred basis is ``sort of" the position basis, i.e., the position basis modulo $a$ (and this choice gives a resolution of the interaction problem).  The superluminal  problem is resolved for, as the authors show, the SL density matrix evolution depends only upon the initial state.  
	
	I thought the SL model was wonderful: I had been stumped on how to resolve the trigger and preferred basis problems and here they were so cleverly demolished.  I also thought it was courageous, because the authors realized that the problems they solved were so important that the work should be published in spite of  problems which were not solved and, indeed, were made more explicit.  With regard to the conservation law violation  problem, the hitting process narrows wave functions and so, by the uncertainty principle, leads to increased energy of particles  (although the values of the parameters $\lambda$ and $a$ were chosen with an eye to keeping that small).  And, the model introduced a new problem, the \textit{symmetry problem}, because the hitting process is not symmetric with respect to all particles, and so destroys the  antisymmetry of the fermionic wave function.  The tails problem is obviously there. 
	
	The SL collapse process is not described by a modified Schr\"odinger equation. The usual unitary Schr\"odinger state vector evolution occurs in between the abrupt SL alterations: the two processes  take place independently, working ``alongside" each other, so to speak. I wondered if it was possible to have them work together.  
		 
\section{Continuous Spontaneous Localization}  
	 
	 It was time for my next sabbatical.  I again wrote to John Bell, on November 13, 1986, asking if it was appropriate to go to CERN or, if not, did he ``... think  that Professors Ghirardi, Rimini or Weber might be interested in inviting me... "  and he replied on November 13:  ``Unfortunately, the CERN Theory Division is more ruthlessly than ever dedicated to the main lines in elementary particle physics.  And most of my own time in the next few years are likely to be devoted to rather practical things concerning accelerator design. I think you would feel isolated and frustrated here (and I also, if you were  here, would be frustrated by not being able to talk with you as much as I would like). So I have written to Ghirardi and Rimini, who are the senior men in that collaboration, and enclosed copies of your letter.  I do not know them personally, but over the years have seen many very sensible papers by them, so that I have a very good opinion of them.  I very much hope that one or the other will be able to invite you, to Trieste or Pavia. "  
	 
	  It was arranged that I spend 3 months in Trieste with GianCarlo  and 1.5 months in Pavia with Alberto. It was, personally and professionally, one of the happiest times of my life.  GianCarlo and his wife Laura, and Alberto and  his wife Silvana, were wonderfully warm and welcoming to myself and my wife Betty.  GianCarlo and I worked hard, but to no avail, on constructing a relativistic SL-type collapse model. But, along the way, I had an idea about how to combine their instantaneous collapse model with my continuous collapse formalism.  Near the end of my stay in Trieste we went to Padua to hear a talk by John Bell, and he was very supportive.  So, when I got to Pavia, Alberto encouraged me to work on my idea.  It was one of those rare golden veins and, in 17 days, I had  written a paper containing what I called the Continuous Spontaneous Localization (CSL) model\cite{PearleCSL}  (which I felt was an appropriate name since it entailed a combination of crucial aspects of my prior work on continuous collapse and the SL model). 
	  
	   In CSL, the  state vector evolution depends upon a fluctuating field $w(\bf{x}, t)$ of white noise type.  There are two equations.  One,  a \textit{linear} Schr\'odinger-type evolution, surprised even me, for it had seemed a truism that the collapse evolution had to be non-linear in the state vector.  The other is an expression for the probability associated with that evolution, and it is this to which 
the nonlinearity is relegated.
	  	  
	  I had couched the CSL model in the language of stochastic differential equations, but it is simpler to understand it, and often to work with it,  if written as follows.  The simplest example, of an initial state vector 
\[
|\psi, 0\rangle=\sum_{n=1}^{N}c_{n}|a_{n}\rangle  
\]	  
(the $|a_{n}\rangle$ are eigenstates of an operator $A$ with nondegenerate eigenvalues $a_{n}$) evolving to one of the  $|a_{n}\rangle$ is 

\begin{eqnarray}\label{11}
|\psi, t\rangle_{w}&\equiv&e^{-(4\lambda)^{-1}\int_{0}^{t}dt'[w(t')-2\lambda A]^{2}}|\psi, 0\rangle\nonumber\\
&=&\sum_{n=1}^{N}c_{n}|a_{n}\rangle e^{-(4\lambda)^{-1}\int_{0}^{t}dt'[w(t')-2\lambda a_{n}]^{2}}.  
\end{eqnarray}
 
 \noindent where $w(t)$ is a random function of white noise type, and $\lambda$ characterizes the collapse rate. The model is comprised of Eq. (\ref{11}) and the rule that the probability associated to $|\psi, t\rangle_{w}$ is
 
 \begin{equation}\label{12}
P_{w}(t)Dw\equiv  _{w}\negthinspace\negthinspace\negthinspace\langle \psi, t|\psi, t\rangle_{w}Dw=\sum_{n=1}^{N}|c_{n}|^{2} e^{-(2\lambda)^{-1}\int_{0}^{t}dt'[w(t')-2\lambda a_{n}]^{2}}Dw  
\end{equation}
\noindent ($Dw\equiv Cdw(0)dw(\Delta t)dw(2\Delta t)...dw(t))$, and $C=(2\pi\lambda/\Delta t)^{-t/\Delta t}$ so that the integrated probability (\ref{12}) is 1).  The density matrix constructed from  (\ref{11}), (\ref{12}) is thus

\begin{equation}\label{13}
\rho=\int P_{w}(t)Dw\frac{|\psi, t\rangle_{w}\thinspace_{w}\langle \psi,t|}{_{w}\langle \psi,t|\psi,t\rangle_{w}}=\sum_{n, m=1}^{N}c_{n}c_{m}^{*}|a_{n}\rangle\langle a_{m} |e^{-(\lambda t/2)(a_{n}-a_{m})^{2}},
\end{equation}
\noindent from which one can see that the off-diagonal elements decay.  This is necessary behavior  if  the individual state vectors which make up the ensemble undergo collapse, but it is not sufficient..  

	Here, without proof\cite{PearleCQC}, is how collapse works for the individual state vectors. If  $T^{-1}\int_{0}^{T}dt'w(t')\rightarrow 2\lambda a_{n}$ as  $T\rightarrow \infty$, then  
\[\lim_{t\rightarrow\infty}\exp-(2\lambda)^{-1}\int_{0}^{t}dt'[w(t')-2\lambda a_{n}]^{2}
\]
 has non-zero measure,  while all the other gaussians asymptotically have vanishing measure. (For any other $w(t)$ behavior, all gaussians have vanishing measure.)  Call a $w(t)$ in this class $w_{n}(t)$.  For a $w_{n}(t)$, Eqs. (\ref{11}) and (\ref{12}) 
become, for large T, 

\begin{equation}\label{14}
|\psi, t\rangle_{w}\approx c_{n}|a_{n}\rangle e^{-(4\lambda)^{-1}\int_{0}^{t}dt'[w(t')-2\lambda a_{n}]^{2}},   
\end{equation}
	  
 \begin{equation}\label{15}
P_{w}(t)Dw\approx |c_{n}|^{2} e^{-(2\lambda)^{-1}\int_{0}^{t}dt'[w(t')-2\lambda a_{n}]^{2}}Dw.    
\end{equation}	  	    
	 
\noindent 	 Thus, (\ref{14}) is a (un-normalized) collapsed state (with a tail, not shown), and the integral of  (\ref{15})'s probability over all $w_{n}(t)$'s is $ |c_{n}|^{2}$ (with a tail contribution).

	For many mutually commuting operators $A_{k}$, and with a possibly time-dependent Hamiltonian $H(t)$ to boot, the evolution (\ref{11}) becomes 

\begin{equation}\label{16}
	|\psi, t\rangle_{w}\equiv {\cal T}e^{-\int_{0}^{t}dt' \{iH(t')+(4\lambda)^{-1}\sum_{k}[w_{k}(t')-2\lambda A_{k}]^{2}\}}|\psi, 0\rangle.  
\end{equation}
\noindent (${\cal T}$ is the time-ordering operator).
	
 	For CSL, I  proposed that the index $k$ correspond to spatial position ${\bf x}$, so that $w_{k}(t)\rightarrow w({\bf x}, t)$ can be regarded as a physical field, and that $A_{k}\rightarrow A({\bf x})$ be the particle number density operator--- which, because of experimental evidence discussed in the companion to this paper, is now to be taken proportional to the mass density operator $M({\bf x})$---``smeared" (using  the SL idea) over a region of length $a$ around $x$:    

\begin{equation}\label{17}
	|\psi, t\rangle_{w}\equiv {\cal T}e^{-\int_{0}^{t}dt' \{iH(t')+(4\lambda)^{-1}\int d {\bf x}[w({\bf x}, t')-2\lambda A({\bf x})]^{2}\}}|\psi, 0\rangle,  
\end{equation}	

\begin{equation}\label{18}
A({\bf x})\equiv\frac{1}{m_{0}(\pi a^{2})^{3/4}}\int  d {\bf z} e^{-\frac{1}{2a^{2}}({\bf x}-{\bf z})^{2}}M({\bf z}),     
\end{equation}
\noindent ($m_{0}$ is taken to be the proton's mass).

	The dynamical equation (\ref{17}) and the probability rule (the first equation in (\ref{12})) constitute the CSL model, which can be applied to any non-relativistic physical system.  CSL works by recognizing a superposition of states which differ in their distribution of mass density, and conducting a gambler's ruin-type competition among them. 
	
		Since operation by $M({\bf x})$, and therefore (\ref{18}), maintains the antisymmetry of a fermionic wave function, CSL solves the symmetry problem (which had been much on my mind because Alberto and GianCarlo were having difficulties resolving it by a modified hitting process\cite{SquiresDove}). Of course, CSL  gives a solution to the interaction problem and, with it, the preferred basis problem since it specifies a particular interaction with collapse toward mass density eigenstates.   And, since the density matrix evolution does not depend upon how the initial density matrix is put together, 
	
\begin{equation}\label{19}
\rho(t)= {\cal T}e^{-\int_{0}^{t}dt' \{iH_{L}(t')-iH_{R}(t')+\frac{\lambda}{2}\int d {\bf x}[A_{L}({\bf x})-A_{R}({\bf x})]^{2}\}}\rho(0) 
\end{equation}
\noindent (the subscripts $L$ and $R$ mean that the operators are to appear to the left or right of $\rho(0)$, and ${\cal T}$ time-reverse orders operators to the right), the superluminal problem is solved.
  
	GianCarlo decided to visited Pavia over the weekend after I showed this work to Alberto and, on Monday, they  had nice things to say about it, and some valuable ideas.  They pointed out to me, since I had not really appreciated it (so focused was I on the physical collapse problem) that I had arrived at a way to obtain the general Lindblad-Kossakowski equation (the most general Markovian density matrix evolution) from a stochastic Schr\"odinger equation, and we decided to write a paper together on the  general framework and its CSL application\cite{GPR}.  Almost near the end of my stay in Pavia, we received a preprint from Gisin\cite{GisinLind} who had independently arrived at a single non-linear stochastic Schr\"odinger equation, which we had shown in \cite{GPR} is equivalent to combining the general framework's linear evolution equation with the non-linear probability rule.   Lajos Diosi\cite{Diosi} and Slava Belavkin\cite{Belavkin}, in modeling continuous observation, also arrived at a similar non-linear formulation, at around the same time.  I later learned from Gisin\cite{GisinGatarek} that, in obtaining the linear Schr\"odinger equation, I had rediscovered something known  to mathematicians as the Girsanov transformation.  	
		
	CSL shows how the Schr\"odinger evolution and the collapse evolution can jointly be described by a single evolution equation.   For the nonlinear collapse process, CSL separates out the non-linear probability rule from the linear evolution. This is a simplification encouraging generalizations, for example, a generalization from the galilean invariant quantum mechanical form of CSL to a relativisitically invariant quantum field theory form.  Because of its linearity, the collapse dynamics (\ref{17}) can directly be expressed in terms of a unitary evolution\cite{Clifton},
\begin{eqnarray}\label{20}
|\psi, t\rangle_{w}&=&\int D\eta e^{-\lambda\int_{0}^{t}dt'd {\bf x}'\eta^{2}({\bf x}',t')}
e^{i\int_{0}^{t}dt'd {\bf x}'\eta({\bf x}',t')w({\bf x}',t')}\nonumber\\
 &&\cdot{\cal T}e^{-i\int_{0}^{t}dt' \{H(t')+2\lambda\int d {\bf x}'\eta({\bf x}', t')A({\bf x}')\}}|\psi, 0\rangle,  
\end{eqnarray}
\noindent (this is just the fourier transform of (\ref{17}), where $\eta$ is the variable conjugate to $w$).  
The last line in (\ref{20}) is the unitary evolution of an initial state vector under the influence of a Hamiltonian consisting of $H$ plus the interaction of a space-time dependent c-number noise field $\eta$ with the smeared mass density operator $A$.  So, according to (\ref{20}), one may view collapse under the field $w$ as a superposition of such unitary evolutions, with a gaussian weight depending upon  $\eta$ and a phase dependent upon $\eta$ and $w$.  
	
	Like SL and, indeed, stimulated by it,  CSL gives a resolution for the interaction, preferred basis, trigger and superluminal problems. It also resolves the symmetry problem raised by SL. In a paper in honor of Abner Shimony, I shall continue discussing the problems remaining in this list, the tails, experimental, conservation law, relativity, and legitimization problems.


\begin{thebibliography}{99}
\bibitem{AbnerFest} I have written in an autobiographical vein once before, P. Pearle, \textit{Experimental Metaphysics: Quantum Mechanical Studies for Abner Shimony}, edited by R. S.. Cohen, M. Horne and J. Stachel (Kluwer, Dordrecht 1997), p. 143.   There is necessarily some anecdotal overlap, which I shall try to minimize.  However,  the present article conceptually stands on its own.  
\bibitem{Bohmtext} D.Bohm, \textit{Quantum Theory} (Prentice Hall, New York, 1951).
\bibitem{Bohmpapers} D. Bohm,  Physical Review  {\bf 85}, 166; 180 (1952).	
\bibitem{Gleason} A. M. Gleason, Journ. Math. Mech.  {\bf 6}, 885 (1957).
\bibitem{PearleAlternative} P. Pearle,  Phys. Rev {\bf 35}, 742 (1967).
\bibitem{Pearleloophole}	 P. Pearle, Am. Journ. Phys. {\bf D2}, 1418 (1970); N. D. Mermin in \textit{New Techniques and Ideas in Quantum Measurement Theory}, (Annals of the New York Academy of Sciences, Vol. 480, New York 1986), p. 422; M. A. Rowe et. al., Nature {\bf 409}, 791 (2001); A. Cabello, Phys. Rev. A, {\bf 72}, 050101 (2005).
\bibitem{BohmBub} D. Bohm and J. Bub, Rev. Mod. Phys. {\bf 38}, 453 (1966).  
\bibitem{Wiener-Siegel} N. Wiener and A. Siegel, Phys. Rev. {\bf 101}, 429 (1956).
\bibitem{Prigogine} I.  Prigogine,  \textit{Non-Equilibrium Statistical Mechanics}, (Interscience, New York 1962). 
\bibitem{Chandrasekhar} S. Chandrasekhar, Revs. Mod. Phys {\bf 15}, 1 (1943), reprinted in N. Wax, 
\textit{Selected Papers on Noise and Stochastic Processes} (Dover, New York, 1954).  
\bibitem{Pearle1} P. Pearle Phys. Rev. D{\bf13}, 857 (1976).
\bibitem{Belinfante} F. Belinfante, \textit{A Survey of Hidden Variable Theories} (Pergamon, Oxford 1973).  
\bibitem{McKean} H. P. McKean Jr., \textit{Stochastic Integrals} (Academic Press, New York 1969).
\bibitem{Wong} E.  Wong, \textit{Stochastic Processes in Information and Dynamical Systems} (McGraw Hill, New York 1972)
\bibitem{Pearle2}P. Pearle, Int'l.  Journ. of Theor. Phys.  {\bf 18},489 (1979).  
 \bibitem{AdlerPercival} See, for example, S. L. Adler, \textit{Quantum Theory as an Emergent Phenomenon} (Cambridge University Press, Cambridge 2004), 
I Percival, \textit{Quantum State Diffusion} (Cambridge University Press, Cambridge 1998), and papers cited in these books.
\bibitem{coins} P. Pearle, Found. of Phys. {\bf 12}, 249 (1982).  
\bibitem{Feller} W. Feller, \textit{An introduction to Probability Theory and its Applications}, (Wiley, New York 1950), Chapter 14.
\bibitem{PearleNYAS} P. Pearle in \textit{New Techniques and Ideas in Quantum Measurement Theory}, D. Greenberger, ed. (New York Academy of Sciences vol. 480, 1986), p. 539.  
\bibitem{Pearle4}  P. Pearle in \textit{The Wave-Particle Dualism}, S. Diner, D. Fargue. G. Lochat and F. Selleri eds. (D. Reidel, Dordrecht 1983), p. 457.   
\bibitem{Penrose} R. Penrose in \textit{Quantum Gravity 2, A Second Oxford Symposium}, C. J. Isham, 
R. Penrose, D. W. Sciama eds. (Clarendon, Oxford 1981), p. 245.
\bibitem{Zeilinger} A. Zeilinger, R. Gaehler, C. G. Shull and W. Treimer, in \textit{Neutron Scattering-1981, Argonne} J. Faber Jr. ed. (AIP, New York, 1982), p. 93.
\bibitem{PearleZeilingerneut} P. Pearle, Phys. Rev. D{\bf 29}, 235 (1984); A. Zeilinger in \textit{Quantum Concepts in Space and Time}, R. Penrose and C. J. Isham eds (Clarendon, Oxford 1986), p. 16.  
\bibitem{Gisin1} N. Gisin, Phys. Rev. Lett. {\bf 52}, 1657 (1984). 
\bibitem{PearlePRLlett} P. Pearle, Phys. Rev. Lett. {\bf 53}, 1775 (1984). 
\bibitem{GisinPRLlett}N. Gisin, Phys. Rev. Lett. {\bf 53}, 1776 (1984).
\bibitem{Pearletime} P. Pearle, Journ. of Stat. Phys. {\bf 41}, 719 (1985).  
\bibitem{PearleSuperlum} P. Pearle,  Phys. Rev. D{\bf 33}, 2240 (1986).
\bibitem{BellGRW} J. S. Bell, in \textit{Schr\"odinger---Centenary Celebration of a Polymath}, C. W. Kilmister, ed. (Cambridge University Press, Cambridge 1987), p. 41.
\bibitem{GRW} G. C. Ghirardi, A. Rimini and T. Weber, Phys. Rev. D{\bf 34}, 470 (1986).
\bibitem{GhirBassi} In the review article, A. Bassi and G. C. Ghirardi, \textit{Physics
Reports} {\bf379}, 257 (2003), Ghirardi presents  ``... a brief review of the historical development of dynamical reduction models.  The history goes back to the years 1970-1973, when G. C. Ghirardi, L. Fonda, A. Rimini and T. Weber were working on quantum decay processes and in particular on the possibility of deriving, within a quantum context, the exponential decay law. Some features of their approach have been extremely relevant for  the subsequent elaboration of the dynamical reduction program," and goes on to specify some features.  I am sure this earlier work was helpful to the authors in constructing the SL model, but it is not part of the history of dynamical reduction. It is part of their intellectual history, just as section 1 of this paper is part of my intellectual history.   Indeed, the 1973 work he cites contains no mention of dynamical reduction (it applies, to decaying particles, the reduction postulate  of the Copenhagen scheme,  brought about as usual by interaction of the particles with an apparatus).  Nor does \cite{GRW} contain any mention of the cited 1973 work.  \cite{GRW} does mention A. Barchielli, L. Lanz and G. M. Prosperi, Nuovo Cimento {\bf 72B}, 79 (1982) and Found. Phys. {\bf 13}, 779 (1983), which introduced a hitting process (a feature adopted in SL)  to model repeated position measurements.  
\bibitem{PearleCSL} P. Pearle, Phys. Rev. A{\bf 39}, 2277 (1989).
\bibitem{PearleCQC} For a recent proof, see P. Pearle, Phys. Rev. A{\bf 72}, 022122 (2005).
\bibitem{SquiresDove} C. Dove and E. J. Squires, Found. Phys {\bf 25}, 1267 (1995) arrived at a hitting process which preserves symmetry. A hit with center {\bf z} involves multiplication by 
$[ \sum_{j}m_{j}\exp-({\bf x_{j}}-{\bf z})^{2}/2a^{2}]^{1/2}$, where $m_{j}$ is the particle's mass. This has also been argued for in R. Tumulka, to be published in Proc.Roy. Soc. A (2006), quant-ph/0508230.   
\bibitem{GPR} G. C. Ghirardi, P. Pearle and A. Rimini,  Phys. Rev. A{\bf 42}, 78 (1990).
\bibitem{GisinLind} N. Gisin, Helv. Phys. Acta {\bf 62}, 363 (1989). Soon after I had left Europe, Alberto and GianCarlo were invited by Nicolas to visit the University of Geneva and GianCarlo gave a seminar on my work and our work together.  Gisin's subsequently published paper differs from his preprint in the concluding section, where it gives CSL model dynamics, and at the end of an earlier section, where generalization to mutually commuting operators $A_{k}$ is given, whereas the preprint dealt with a single operator $A$.  
\bibitem{Diosi} L. Diosi, Physics Letters A {\bf 132}, 233 (1988).
 \bibitem{Belavkin} V. P. Belavkin, A {\bf 149}, 355 (1989).
\bibitem{GisinGatarek} D. Gatarek, N. Gisin, Journ. Math. Phys. {\bf 32}, 2152 (1991).
\bibitem{Clifton} P. Pearle in \textit{Perspectives on Quantum Reality}, R. Clifton ed. (Kluwer, Dordrecht 1996), p. 93.
\end{thebibliography}
\end{document}